\documentclass[10pt, twocolumn]{article}
\usepackage{times}
\usepackage{setspace}
\usepackage{amsmath,amssymb,amsfonts} % Typical maths resource packages
\usepackage{subeqnarray}
\usepackage{empheq}
\usepackage{graphicx}                 % Packages to allow inclusion of graphics
\usepackage{color}                    % For creating coloured text and background
\usepackage{hyperref}                 % For creating hyperlinks in cross references
\usepackage{makeidx}
\usepackage{makecell}    
\usepackage[backend=bibtex, style=numeric-comp, natbib=true, sorting=none, maxbibnames=99]{biblatex}
\usepackage{fancyhdr}
\renewbibmacro{in:}{}
\usepackage{orcidlink}

\DeclareGraphicsExtensions{.jpg}

\makeindex

{\end{quote}}

\title{Equations of motion for dust particles in dust-collecting devices {\normalsize \Large (Uravneniya dvizheniya pylevykh chastits v pylepriyemnykh ustroystvakh)}} 

\author
{L. S. Klyachko, Engineer \\
\\
\\
}

\date{}

\fancypagestyle{firstpage}{
    \fancyhf{} % Clear first page settings
    \fancyhead[L]{Translated by Ramírez, Feng, and Kubesh from USSR Journal {\em Otoplenie i Ventilyatsiya}, No. 4 (1934): 27-29}
    \fancyhead[R]{ }
    \fancyfoot[C]{\thepage}
}

\begin{document} 

% --- Front Page ---
\begin{titlepage}
%    \maketitle
\noindent
{\Large \bf English translation of ``Uravneniya dvizheniya pylevykh chastits v pylepriyemnykh ustroystvakh'' by L. S. Klyachko, originally published in USSR Journal {\em Otoplenie i Ventilyatsiya}, No. 4 (1934): 27-29. This translation by C. A. Ramírez, J. Q. Feng, and R. J. Kubesh is distributed under an arXiv non-exclusive license. The original Soviet work has entered the public domain worldwide, as clarified by the Russian Library in Moscow regarding the copyright of this Klyachko article.}

\hspace{15pt}

\hspace{15pt}

\hspace{15pt}

\section*{\large Preface for the English Translation}
{\large While checking out early USSR journal articles for our literature search, we recently had tremendous difficulty acquiring this 1934 Klyachko original publication through our library systems. After obtaining and reading a scanned copy of it supplied by the Russian library in Moscow, we realized that relatively scarce citations of this work in the Western literature are often accompanied by inaccurate descriptions and comments, indicating that most Western authors might have gotten only second-hand knowledge about it and never seemed to have read the original article.  Compared with various correlation formulas in the modern literature, the Klyachko 1934 formula for sphere drag in fluid flow appears to be the most elegant and cleanest one thus far. Therefore, we decided to make an effort to translate this historical work and publish it in the open literature to facilitate its accessibility to English readers worldwide.
}

\hspace{15pt}

\hspace{15pt}

\noindent
C.A.R.

\noindent
J.Q.F.

\noindent
R.J.K.

    \thispagestyle{empty} % Removes page number from the title page
\end{titlepage}

% --- Main Content ---
\clearpage
\pagenumbering{arabic} % Starts numbering from 1 on this page

\renewcommand{\thefootnote}{\fnsymbol{footnote}}


%\baselineskip24pt


\maketitle 


%\begin{abstract}
%\noindent 
%\end{abstract}

\thispagestyle{firstpage}

%\noindent \textbf{Keywords:}
%\begin{multicols}{2}

Issues Nos. 2, 3, and 4 of the journal ``Heating and Ventilation'' (1933) published articles by Engineer Vavakin concerning the derivation of the so-called fundamental equation of motion for dust particles; this derivation was based on an approximate expression he provided for $\varphi = f(Re)$, where $Re = w \cdot d / \nu$ is the Reynolds number and $\varphi$ is the drag coefficient (in the aforementioned article by Vavakin, the corresponding notations used were $P$ and $K$). 

Without addressing in this article the question of whether this functional relationship alone suffices to solve the problem he posed—a matter we consider highly doubtful, given that the condition is necessary but not sufficient—we shall focus solely on the functional relationship itself as presented by Vavakin.

The main drawback of this relationship is its discontinuity and the consequent need to employ a series of intermediate solutions for the range 
\[
0 < Re \le 1000 \, .
\]

In view of this, the specific expression for $\varphi$ proposed by the author lacks physical significance, as it fails to yield the Stokes expression for $\varphi$ as $Re \to 0$.

In 1929, while investigating a number of issues in the field under consideration and in related fields, we obtained the following expression for the quantity $\varphi = f(Re)$ for spherical particles: 
 \begin{equation}
\varphi = 
  \frac{24}{Re} + \frac{4}{Re^{1/3}} \quad \, 
\footnote{This expression for $\varphi$ was reported by the author at a meeting of the Scientific and Technical Society (NTO) of the Teplomontazh L/o local group in November 1929; as far as known to the author, the same definition for $\varphi$ is given in one of the articles in the journal “Rauch und Staub” for 1930 and also in one of the volumes of the “Handbuch der Physik” of 1932.}
  \label{klyachko}
\end{equation}      
where $24/Re$ is the Stokes value of the drag coefficient $\varphi$, as can be readily verified from the following transformations. As is well known, Stokes arrived at the following expression for the drag on spherical particles using a purely analytical approach: 
 \begin{equation}
 R = 6 \pi r \mu w = 3 \pi d \mu w 
  \,\,\,  ,
  \label{stokes}
\end{equation}
where $R$ is the drag force on particle motion in kg; $r$ is the radius of the particle (sphere) in m; $d$ is the diameter of the particle (sphere) in m; $\mu$ is the dynamic viscosity of the surrounding fluid in kg·s/m²; and $w$ is the velocity of the particle in m/s.

Expression (\ref{stokes}) can be written in the form:
 \begin{equation}
 R = \varphi \frac{ \pi d^2}{4} \frac{w^2 \gamma}{2 g}  = 3 \pi d \mu w 
  \,\,\,  ,
  \label{R_def}
\end{equation}
where
 \begin{equation}
 \varphi = 3 \cdot 2 \cdot 4 \cdot \frac{ g \mu}{d w \gamma} = \frac{24}{Re}
  \,\,\,  ,
  \label{varphi_def}
\end{equation}
where $\gamma$ is the specific weight of the medium surrounding the sphere, and $g$ is the acceleration due to gravity in m/s².  \footnote{When comparing the drag coefficients given in this article by engineer L. S. Klyachko and in the above-mentioned articles by engineer T. S. Vavakin, it should be borne in mind that the former uses the drag formula in the form $R = \varphi \rho F v^2/2$, and the second in the form $R = K \rho F v^2$, so there is a factor of $\varphi /K = 2$ difference. Taking the formula in the second form, we obtain Stokes’ law (for a smooth sphere):  $K \rho F v^2 = 3 \pi \mu d v$ whence: $K = 12/Re$. }

Regarding the derivation itself as applied by Stokes, this expression for $\varphi$ is valid only for attached flow around the sphere—a condition that, logically speaking, is possible only at Reynolds number close to zero (i.e., as $Re \to 0$). 

Given the limitations of the Stokes formula, a number of researchers have adopted new analytical derivations and conducted purely experimental investigations of the problem, because relying solely on mathematics in this instance is unreliable and inadequate.

Indeed, once the Swedish scientist  Oseen identified the cause of the shortcomings in Stokes's solution—namely, the neglect of inertial forces—and introduced just one additional term into the general differential equation of fluid motion to account for them, the solution itself became extremely complex; yet, the result was merely a particular value for $\varphi$ of the form:
\begin{equation}
\varphi = 
  \frac{24}{Re} \left(1 + \frac{3}{16}{Re} \right) \, 
  \,\,\,  ,
  \label{oseen}
\end{equation}
which slightly extended the range of values of $f(Re)$.

Without dwelling further on the theoretical solutions proposed by other scientists, we shall point to the purely experimental solution obtained at the Göttingen laboratory by Prandtl’s students, as well as to the independent study conducted in the USSR by Prof. Velikanov (jointly with Zegzhda, 1930). The results of the Göttingen experiments are presented as a curve $\varphi = f(Re)$, whereas Velikanov’s experiments yielded a highly complex transcendental function requiring a graphical solution.

It is evident that neither result is applicable to the analytical treatment of the problem—where one must use the expression for $\varphi$ as a continuous function of $Re$ over a wide range of Reynolds number, as is the case here or in other problems such as the heat transfer from a moving sphere in a medium.

In 1929, based on Osborne Reynolds' concept and a body of experimental data regarding heat transfer, we established the need to introduce an additional term into the expression $\varphi = f(Re)$, with an exponent for $Re$ close to $-1/3$. By subsequently comparing the general form of the expression we derived for $\varphi$ with the aforementioned curve provided by the Göttingen laboratory, we determined the value of the constant for the second term—namely, $4$—and thus obtained the final expression for $\varphi$ in the form (\ref{klyachko}):
 \[
\varphi = 
  \frac{24}{Re} + \frac{4}{Re^{1/3}} \quad  \, . \quad \quad \quad (1)
\] 

Incidentally, the expression for the heat transfer coefficient that we obtained has the form:
 \begin{equation}
Nu = 
 C (A + B {Re^{2/3}})  \, ,  \,\, \footnote{We will discuss this issue further in a special article.  This formula was proposed in 1929 by the author of this work for discussion by LOTI staff, in particular Profs. A. A. Gukhman and V. S. Zhukovsky.  I take this opportunity to express my gratitude to them for their suggestions.}
  \label{Nu_def}
\end{equation}
where $C$, $A$, and $B$ are constants for $0 < Re \le 1000$. 

Having thus determined the method for establishing the expression for $\varphi$, we proceed directly to the question that interests us.

We previously utilized formula (\ref{klyachko}) in our 1931 study on pneumatic conveying, where, to simplify the solution, we adopted an integral mean value for $\varphi$ in the specific example presented there. The advantage of this functional relationship lies in its continuity and applicability across the entire range of problems related to dust transport ($0 < Re \le 1000$). 

However, for $Re$ values in the range from $1000$ up to Prandtl’s drag crisis ($1000 \le Re \le \approx 3 \cdot 10^5$), $\varphi$ can be assumed—with sufficient accuracy—to be constant and equal to the value of $\varphi$ corresponding to $Re = 1000$.

Thus, in accordance with the objective of this study, we are to solve the following differential equation:
 \begin{equation}
 m \frac{d^2 S}{dt^2} = 
  - \left(\frac{24}{Re} + \frac{4}{Re^{1/3}}\right) \frac{\pi d^2}{4} \frac{\rho w^2}{2} \quad  \, ,
  \label{particleAccel}
\end{equation}
where $m$ is the mass of the moving particle (sphere); $d$ is the diameter of the sphere; $\rho$ is the volumetric density of the medium surrounding the sphere; $S$ is the distance traveled by the particle; and $w$ is the particle velocity ($dS/dt$).

Omitting the derivation details here due to their complexity and the cumbersome nature of the transformations involved, we merely note that, after expanding the expression for $\varphi$ and integrating within the limits from $Re_{1}$ to $Re_{2}$—corresponding to $t$ from $0$ to $\tau$ and for $S$ from $0$ to $S$—we obtain the following final expressions:
  \begin{equation}
 \tau = \frac{\rho_p d^2 }{12 \mu} 
 \ln \left[\frac{Re_1^{2/3} (Re_2^{2/3} + 6)}{Re_2^{2/3} (Re_1^{2/3} + 6)}\right] \quad  \, ,
  \label{particle-tau}
\end{equation}

 \begin{equation}
\begin{array}{c} S = \sqrt{6} \frac{\rho_p d}{\rho} \left[\sqrt{\frac{Re_1^{2/3}}{6}} + \tan^{-1}\left(\sqrt{\frac{6}{Re_1^{2/3}}}\right)  \right .\\ \\
  - \dfrac{1}{\sqrt{\left(1 + \frac{6}{Re_2^{2/3}}\right) \exp \left(\frac{12 \mu}{\rho_p d^2} \tau \right) - 1}}   \quad  \\  \\
 \left .  - \tan^{-1} \left( {\sqrt{\left(1 + \tfrac{6}{Re_2^{2/3}}\right) \exp \left(\tfrac{12 \mu}{\rho_p d^2} \tau \right) - 1}} \right) \right]  \, ,  
    \end{array}
  \,  
  \label{particle-S}
\end{equation}     
where $\tau$ is the flight duration; $Re_1$ is the initial $Re$ value corresponding to the initial particle launch velocity $w_1$ (and its diameter); $Re_2$ is the final $Re$ value (usually the one to be determined); and $\rho_p$ is the bulk density of the particle material. 

It now remains to consider the question of the terminal settling velocity $w_s$ or the Reynolds number ${Re}_s$, respectively.

In this part, the shortcomings of Vavakin's functional relationship—specifically its discontinuity—are especially evident.

Using formula (\ref{klyachko}), we obtain the following expression for the steady-state regime:
  \begin{equation}
 P_s = \frac{\pi d^3}{6} {\gamma_s}
 =\left(\frac{24} {Re_s} + \frac{4}{Re_s^{1/3}} \right) \frac{\pi d^2}{4} \frac{\rho w_s^2}{2}  \quad  \, ,
  \label{P_def}
\end{equation}
where $P_s$ is the weight of the sphere; $\gamma_s$ is the specific weight of the sphere material; $Re_s$ is the corresponding $Re$ value for velocity $w_s$.

Expanding the expression for $Re$ and performing the appropriate transformations, we finally obtain:
 \begin{equation}
 P_s = 3.14 \frac{\mu^2}{\rho}
 \left(3 {Re_s} + 0.5 {Re_s^{5/3}} \right) \quad  \, ,
  \label{P_p}
\end{equation}
for $0 < Re \le 1000$. 

From the last formula (\ref{P_p}), it is easy to determine the value of the terminal settling velocity $w_s$: 
 \begin{equation}
 w_s = \frac{Re_s \nu}{d}
 \quad  \, ,
  \label{w_p}
\end{equation}
where $\nu$ is the kinematic viscosity of the surrounding fluid; $d$ is the particle (sphere) diameter.

Thus, the expressions we have presented for $\varphi$, $Re$, $\tau$, and $S$ are generalized, since the function $\varphi = f(Re)$ is continuous over the entire range $0 < Re \le 1000$; it not only fully replaces all the specific solutions cited in Vavakin’s article but also naturally yields more accurate results.

In conclusion, we consider it necessary to note that the derivation of the basic equation of motion for dust particles, which is the simplest equation of the dynamics of a point mass, was given in great detail in our earlier work on pneumatic transport\footnote{1928/29 edition, and 1931 second edition}, where the integral mean value is adopted for $\varphi$ (as for a function of a single variable) to simplify the solution of the specific problem, as noted in the footnote in that work.
 
%\end{multicols}

\begin{figure}[ht] \label{fig1}
\includegraphics[clip=true,scale=0.62,viewport=180 30 580 510]{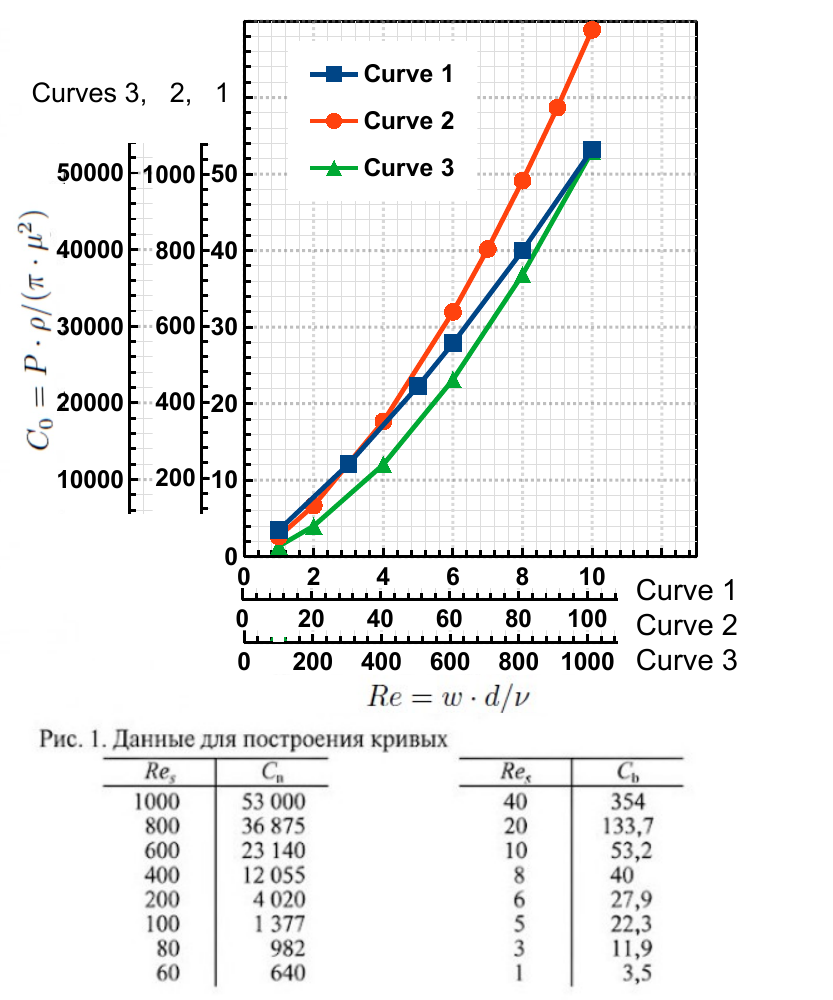}
\caption{Graphic curves of $C_0$ ($= 3 Re + 0.5 Re^{5/3}$) versus $Re$
% (with curves 1, 2, 3 for three different value ranges of $Re$)
, and tabulated data for plotting the curves. }
\end{figure}

%\begin{multicols}{2}

To clarify the use of formulas (\ref{particle-tau}), (\ref{particle-S}), and (\ref{P_p}), we present a numerical example.

Given: particle diameter $d = 10^{-4}$ m; initial velocity $w_1 = 26$ m/s; surrounding medium — air at $T = 0^o$C and $p = 760$ mm Hg. The dynamic viscosity of the medium: $\mu = 1.7 \cdot 10^{-6}$ kg·s/m$^2$; kinematic viscosity $\nu = \mu / \rho = 13 \cdot 10^{-6}$ m$^2$/s; density of the medium $\rho = 0.131$ kg·s$^2$/m$^4$; specific weight of the sphere material $\gamma = 2 \cdot 10^3$ kg/m$^3$.

Determine the flight time $\tau$ and the distance traveled $S$, for the final velocity $w_2 = 1.3$ m/s. 
Calculate $Re_1$ and $Re_2$:
\[
Re_1 = \tfrac{w_1 \cdot d}{\nu} = \tfrac{26 \cdot 10^{-4}}{13 \cdot 10^{-6}} = 200 \,\, ,
\]
\[
Re_2 = \tfrac{1.3 \cdot 10^{-4}}{13 \cdot 10^{-6}} = 10 \,\, ,
\]
then using formula (\ref{particle-tau}), we obtain:
\[
\begin{array}{c}
\tau = \frac{2000 \cdot 10^{-8} }{12 \cdot 9.81 \cdot 1.7 \cdot 10^{-6}} 
\ln \left[\frac{200^{2/3} \cdot (10^{2/3} + 6)}{(200^{2/3} + 6) \cdot 10^{2/3}} \right] \\
= 0.0668 \, \mathrm{sec} \,\, . \qquad \qquad \qquad \qquad \qquad
\end{array}
\]

Accordingly, the distance traveled will be, from formula (\ref{particle-S}):
\[
\begin{array}{c}
S = \sqrt{6} \frac{2000 \cdot 10^{-4} }{9.81 \cdot 0.131} \left[\sqrt{\frac{34}{6}} + \tan^{-1}\left(\sqrt{\frac{6}{34}}\right)   \right .\\ \\
  - \dfrac{1}{\sqrt{\left(1 + \frac{6}{4.64}\right) \exp \left(\frac{12 \cdot 9.81 \cdot 1.7 \cdot 10^{-6}}{2 \cdot 10^{-5}} \cdot 0.0668 \right) - 1}}  \quad  \\  \\
 \left .  - \tan^{-1} \left( {\sqrt{\left(1 + \tfrac{6}{4.64}\right) \exp \left(0.668 \right) - 1}} \right) \right] 
 \approx 0.44 \, \mathrm{m}  \, .  
\end{array}
\]
(Some error must have occurred to arrive at "0.32 m" in this calculation in Klyachko's original work.) 

The suspension velocity for this same particle can be determined from equation (\ref{klyachko}), since we know that:
\[
P_s = \frac{\pi d^3 \gamma_s}{6} \, .
\]

Then, according to formula (\ref{P_p}), we have
\[
\begin{array}{c}
3 Re_s + 0.5 Re_s^{5/3} = \dfrac{\pi d^3 \gamma_s \rho}{6 \pi \mu^2} = \\ \\
= \frac{10^{-12} \cdot 2 \cdot 10^3 \cdot 0.131}{6 \cdot 1.7^2 \cdot 10^{-12}} = 15.1
\end{array}
\]
where
\[
Re_s \approx 3.6 \, \, .
\]

The value from the last equation $Re_s$ can be determined either by the trial-and-error method commonly used in engineering practice or by means of the graph in Fig. 1, which is compiled for the values of $Re$ from $0$ to $1000$ according to the formula: 
\[
3 Re_s + 0.5 Re_s^{5/3} =C_0 \,\, ,
\]
with
\[
C_0 = \frac{P_s \cdot \rho}{3.14 \cdot \mu^2} \,\, .
\]

The values $C_0$ are plotted along the y-axis, and the values $Re_s$---on the x-axis.

For the present particular problem, we have:
\[
Re_s = 3.6 \, \, ,
\]
then obtain
\[
w_s = \frac{ 3.6 \cdot 13 \cdot 10^{-6}}{10^{-4}} = 0.468 \,  \mathrm{m/s} \,\, .
\]

\hspace{15pt}

%\hspace{15pt}

\noindent
{\em Leningrad }
%\end{multicols}

\noindent\hrulefill

\hspace{15pt}

\section*{\normalsize Afterword for the English Translation}
In 1934, Soviet engineer Lev Solomonovich Klyachko appears to be the first to publish the drag coefficient formula for a sphere moving in an incompressible Newtonian fluid (in an article entitled ``Uravneniya dvizheniya pylevykh chastits v pylepriyemnykh ustroystvakh''
in USSR Journal {\em Otoplenie i Ventilyatsiya}, No. 4 (1934): 27-29), as
 \[
\varphi = f(Re) =  \left\{  \begin{array}{l} 
  \frac{24}{Re}\left(1 + \frac{Re^{2/3}}{6}  \right) , \mbox{\small for 0 } {\tiny < Re \le} \mbox{\small1000} \\  \\
   0.424  \, , \mbox{\small for 1000 } {\normalsize \tiny < Re < 3 \times 10^5} \,\, ,
    \end{array}\right .  
  \,
\] 
(which was substantially established in 1929, according to Klyachko's mention in the article), reasonably accurate for most practical applications. 

However, this 1934 Russian paper (though in the public domain globally) has been largely omitted from citations in the literature, probably because it is not easily accessible
to English readers, as we experienced with our library systems.  Many authors, although they attempted to cite Klyachko's
formula ($\varphi = 24/Re + 4/Re^{1/3}$), never seemed to have read the original article.
After a long, tedious search through the Puerto Rico library system and several international libraries, we acquired a scanned copy from the Central Scientific and Technical Library in Moscow.  With a combined effort using {\em Google Translate}, {\em Copilot}, and old-fashioned typing on a Russian keyboard for specific word translations to ensure technical accuracy, we completed the English translation of this historical work. We are now happy to share it with the scientific community. 

It should be noted that USSR engineers in 1934 were using kg as the unit of force; as a consequence, the values of force, dynamic viscosity, and density in the Klyachko article (presented in obsolete metric units) should be multiplied by $g = 9.81$ m/s$^2$ to convert them to the corresponding modern SI units.

Herewith, we would like to acknowledge the tremendous efforts of librarians in Puerto Rico and Moscow in locating and delivering a scanned copy of the original Klyachko 1934 article.

\hspace{15pt}

%\hspace{15pt}

%\renewcommand{\thefootnote}{\fnsymbol{footnote}}

%\begin{doublespace}
\noindent
{Carlos A. Ramírez}\footnote{University of Puerto Rico, Mayagüez, Puerto Rico} ( \orcidlink{0000-0002-8148-3985} 0000-0002-8148-3985 )

\noindent
{James Q. Feng}\footnote{OXCBO Research, Maple Grove, Minnesota, USA}  ( \orcidlink{0000-0003-4041-3179} 0000-0003-4041-3179 )

\noindent
{Rodney J. Kubesh}\footnote{St. Cloud State University, St. Cloud, Minnesota, USA} ( \orcidlink{0009-0001-7754-1724} 0009-0001-7754-1724 )

%\end{doublespace}

%$C_0 =P \cdot \rho / (\pi \cdot \mu^2)$ 

%\hspace{15pt}

%$Re = w \cdot d /\nu$

%Fig. 1. Data for plotting the curves

\end{document}